\documentclass[runningheads]{llncs}
\usepackage[T1]{fontenc}
\usepackage{graphicx}
\usepackage{amssymb}
\usepackage{booktabs}
\usepackage{tcolorbox}
\usepackage{multirow}
\begin{document}
\title{Reading between the Lines: Leveraging Large Language Models for Global Dementia and Depression Assessment from Clinical Interviews}
\titlerunning{LLMs for Global Dementia and Depression Assessment}
\author{Franziska Braun\inst{1}\orcidID{0000-0002-4670-3807} \and
Alea R\"uggeberg\inst{2} \and
Thomas Ranzenberger\inst{1} \and
Hartmut Lehfeld\inst{3} \and
Thomas Hillemacher\inst{3} \and
Tobias Bocklet\inst{1} \and
Korbinian Riedhammer\inst{1}}
\authorrunning{F. Braun et al.}
\institute{\textsuperscript{1}TH N\"urnberg, \textsuperscript{2}FAU Erlangen, \textsuperscript{3}PMU Klinikum N\"urnberg, Germany\\
\email{franziska.braun@th-nuernberg.de}}
\maketitle              
\begin{abstract}
Dementia and depression are the most prevalent neuropsychiatric disorders in geriatric populations, and their overlapping symptoms pose major challenges for differential diagnosis.
In this study, we investigate open-weights Large Language Models (LLMs) for predicting dementia and depression severity from speech samples collected during standardized history taking interviews with 154 German-speaking subjects.
We introduce an observer-based Global Depression Scale (GDS-D) aligned with the established Global Deterioration Scale (GDS), enabling parallel global staging of affective and cognitive symptoms.
We compare three LLMs (Mistral~3.1, DeepHermes, Qwen3) in two settings: (1) zero-shot prediction and (2) LLM-based feature extraction for Support Vector Regression, using human and pause-enriched transcripts.
Results show that LLMs effectively predict depression severity in zero-shot settings (best MAE of 0.60), while dementia assessment benefits substantially from structured feature extraction (best MAE of 0.78), reducing errors by up to 35\% over zero-shot baselines.
Pause-enriched transcripts achieve competitive performance with human transcriptions, demonstrating the viability of fully automatic screening pipelines for differential neuropsychiatric assessment.

\end{abstract}
\section{Introduction}

Dementia and depression are the most prevalent neuropsychiatric disorders in older adults and pose major challenges for differential diagnosis.
Both conditions frequently present with overlapping cognitive and affective symptoms, complicating the distinction between pseudodementia and neurodegenerative cognitive decline.
Gold standard differential diagnostics therefore rely on history taking, neuropsychological testing, biomarkers, and questionnaires.
Recent advances in Large Language Models (LLMs) offer a cost-effective, low-barrier tool for automatic screening from unstructured clinical text and speech data, with the potential to support diagnostics and reduce the burden on healthcare systems.
Pretrained LLMs have demonstrated strong performance across diverse Natural Language Processing (NLP) tasks, including dementia and depression screening, without task-specific fine-tuning; however, nuanced clinical assessment, particularly involving geriatric depression, remains underexplored, limiting applicability for differential diagnostics.
In this work, we investigate open-weights LLMs for global dementia and depression assessment using speech collected during clinical history taking, which typically constitutes the first step in routine evaluations and informs subsequent diagnostic decisions.
We introduce a symptom-based Global Depression Scale aligned with the Global Deterioration Scale for dementia staging, enabling parallel staging of affective and cognitive symptoms.
Our baseline experiments evaluate LLM zero-shot prediction of both scales using human and pause-enriched automatic transcriptions.
In an advanced setting, LLMs serve as feature extractors for transcript-derived feature sets (reported symptoms, behavioral \& psychopathologic observations, structural markers, and handcrafted language features), which are used to train Support Vector Regression (SVR) models.

\section{Related Work}
Dementia assessment typically relies on standardized scales such as the Mini-Mental State Examination (MMSE) \cite{mmse75}, the Montreal Cognitive Assessment (MoCA) \cite{moca_2005}, and the Global Deterioration Scale (GDS) \cite{reisberg_global_1982}.
Speech-based dementia detection has primarily focused on features extracted from cognitive tasks such as picture descriptions (Pitt Corpus), targeting MMSE prediction, Alzheimer's disease (AD) classification \cite{adress20,adresso21,madress22,taukadial24}, or GDS-based staging \cite{braun_GoingCookieTheft_2022,braun24_interspeech}, while depression discrimination from cognitive tasks remains limited \cite{braun_classifying_2023}.
LLM-based approaches employ zero-shot prediction, feature extraction, and fine-tuning for AD classification \cite{agbavor2022lldemcookie,botelho24_interspeech}, though they often rely on binary labels.
Spontaneous speech from clinical interviews remains understudied for cognitive impairment detection \cite{braun_GoingCookieTheft_2022}.

Depression severity is commonly assessed using scales such as the Beck Depression Inventory (BDI) \cite{beck1961}, Hamilton Depression Rating Scale (HAM-D) \cite{hamilton56}, Patient Health Questionnaire-9 (PHQ-9) \cite{phq92001}, and Montgomery-\AA sberg Depression Rating Scale (MADRS) \cite{montgomery1979}, which are designed as symptom sum scores or rating-based instruments rather than global staging models.
Speech-based studies predominantly use PHQ-9 labels derived from clinical interviews (DAIC-WOZ benchmark) \cite{avec2016,avec2019}.
Most LLM-based approaches operate on plain text data such as social media posts \cite{wang2024llmdepsocial}, diaries \cite{shin2024llmdepdiary}, narratives \cite{jama2025llmdepnarratives}, and clinical notes \cite{thomas2025llmdepnotes}, with only few studies using spoken interview transcripts \cite{xin2024llmdepinterview}.
While promising, these studies typically involve average-aged adults, overlooking geriatric depression despite its distinct symptomatology and affective profiles (e.g., pseudodementia).

To address these limitations, we investigate LLMs for nuanced dementia and depression assessment in older adults, introducing an observer-based Global Depression Scale (GDS-D) aligned with the GDS for comorbidity modeling through parallel staging of affective and cognitive symptomatology.

\section{Data}\label{sc:data}

We use a subset of the corpus introduced in our ongoing study in 2022 \cite{braun22_interspeech}, comprising 154 German-speaking subjects (62 men, 92 women) aged 49 to 89 years ($\mu=74.05\pm8.78$).
All interviews and corresponding recordings were collected during routine clinical practice as part of a face-to-face dementia screening procedure, including history taking, cognitive testing, and questionnaires.
Notably, the healthy control group was drawn from the same clinical pool rather than recruited separately, reducing selection bias.
Expert diagnoses ranged from non-impaired to moderate stages for both dementia and depression, informed by medical examinations, questionnaires, cognitive testing, and history taking.
Some subjects exhibit strong regional accents and dialects, and all speakers wore surgical masks during recording, posing additional challenges for speech processing.
\begin{table}[!th]
\centering
\caption{\label{tab:infr}History-taking interview: reported symptoms (S) and behavioral \& psychopathological observations (O).}
\begin{tabular}{c||l|l}
\midrule
\textbf{S} & \textbf{category} & \textbf{question}                       \\
\midrule
\textbf{1} & Memory             & Does the person notice memory problems?         \\
\textbf{2} & Attention          & Does the person notice attention problems?      \\
\textbf{3} & Daily living       & Is the person dependent on help? \\
\textbf{4} & Orientation        & Is the person oriented in time and space?       \\
\textbf{5} & Mood               & Is the person's mood positive or negative?      \\
\textbf{6} & Sleep              & How is the person's sleep quality?              \\
\textbf{7} & Appetite           & How is the person's appetite?                   \\
\textbf{8} & Physical activity  & Is the person physically active or does sports?\\
\midrule
\midrule
\textbf{O}  & \textbf{category} & \textbf{question}                                                 \\
\midrule
\textbf{1}  & Signs of Fatigue   & Does the person appear tired?                                             \\
\textbf{2}  & Aphasia            & Are there signs of aphasia, e.g. ...? \\
\textbf{3}  & Tangentiality      & Does the person tend to stray off-topic?                                  \\
\textbf{4}  & Distractibility    & Does the person appear easily distracted?                                 \\
\textbf{5}  & Perseveration      & Does the person stick to or repeat thoughts?                     \\
\textbf{6}  & Complaining        & Does the person appear negative/complaining?                             \\
\textbf{7}  & Disinhibition      & Are there inappropriate/intrusive tendencies?                          \\
\textbf{8}  & Agitation          & Does the person appear restless?                                          \\
\textbf{9}  & Rejection          & Does the person have a rejecting attitude?                                \\
\textbf{10} & Aggressiveness     & Are there aggressive tendencies?                                         \\
\midrule
\end{tabular}
\end{table}
We use speech samples from patient--psychologist interactions during history taking interviews, ranging in duration from 5.8 to 33.7 minutes.
Each interview follows a standardized psychological assessment template (cf.\ Tab.~\ref{tab:infr}), comprising nine open-ended questions covering cognitive and affective symptom domains, as well as checkboxes for behavioral and psychopathological observations.

\subsection{Global Deterioration Scale}\label{sc:gds}
Reisberg's Global Deterioration Scale (GDS) \cite{reisberg_global_1982} is a widely used observer-based staging model that assesses cognitive decline associated with primary degenerative dementia across seven stages, each characterized by specific cognitive, functional, and behavioral features.
Stages 1 to 3 cover the spectrum from normal cognitive function to mild cognitive impairment (MCI), while stages 4 to 7 reflect increasing dementia severity (details in \cite{reisberg_global_1982}).

\subsection{Global Depression Scale}\label{sc:gdsd}
We introduce an observer-based Global Depression Scale (GDS-D) aligned with the GDS to enable systematic evaluation of depression symptomatology in geriatric patients.
The GDS-D classifies depressive symptom severity across seven stages, with thresholds and descriptive categories closely modeled after the GDS to ensure conceptual alignment.
This parallel structure facilitates integrated, two-dimensional evaluation of dementia and depression severity for differential diagnostics.
The seven GDS-D stages are summarized in Tab.~\ref{tab:gdsd}.
The staging criteria, conceptualized as an interval-scaled approximation of depression severity, draw on the Geriatric Depression Scale (GDS-K \cite{gdsk2011}) and ICD-10 criteria for depressive episodes \cite{icd102025}.
Key dimensions include the severity and persistence of depressive symptoms, functional impact on daily living, psychomotor retardation, and suicidal ideation.

\begin{table}[!th]
\centering
\caption{\label{tab:gdsd}Global Depression Scale (GDS-D): seven stages (St.) of depression severity.}
\begin{tabular}{c||l|p{8.2cm}}
\toprule
\textbf{St.} & \textbf{Severity} & \textbf{Description} \\
\midrule
\textbf{1} & No depression & Completely healthy, positive outlook. No lack of drive or distress. \\
\textbf{2} & Very mild & Generally well, occasional minor mood fluctuations or mild lack of drive. \\
\textbf{3} & Mild & Mild psychological strain (low mood, worries); slight lack of drive without significant daily impairment. \\
\textbf{4} & Moderate & Regular strain diminishing joy; lack of drive affects daily activities. \\
\textbf{5} & Moderately severe & Severe strain strongly impairing daily life; pronounced lack of drive. \\
\textbf{6} & Severe & Severe depression, intense lack of drive; self-care barely possible. Passive death wishes. \\
\textbf{7} & Very severe & Severe depression with intense lack of drive; suicidal thoughts or actions. Self-care barely possible. \\
\bottomrule
\end{tabular}
\end{table}
Experts assigned a stage per subject by aligning GDS-D criteria with quantitative GDS-K scores and qualitative analysis of personal and collateral history.
Acute grief reactions were categorized identically to persistent depressive states to capture the patient’s current affective status.

\section{Method}
\subsection{Human and Automatic Transcriptions}\label{sc:trans}
Human ground truth (GT) transcriptions were created by a professional service following the extended scientific transcription rules of Dresing \& Pehl \cite{dresing_pehl10}.
These include markers for word and sentence breaks, pauses, filler words, emphasis, non-verbal utterances, speaker changes and overlaps, guiding the conversational structure.

For automatic speech recognition (ASR), we employ the faster-whisper conversion of OpenAI's whisper-large-v3 model \cite{whisper} with the CTranslate2 inference engine.
Model weights are openly available\footnote{\url{https://hf.co/Systran/faster-whisper-large-v3}}.
Complete audio files are transcribed to German using default parameters and Voice Activity Detection (VAD) to obtain accurate word-level timestamps.
We create pause-enriched transcripts that provide temporal context for evaluating speech fluency and cognitive slowing, both of which have been shown to be informative for dementia and depression screening \cite{yuan_disfluencies_2020,braun_classifying_2023,liu2017speechpause}.
Pause durations are computed from word-level timestamps, rounded to the nearest second, and inserted into the transcripts in parentheses.

\subsection{Large Language Models}\label{sc:llm}
We compare three open-weights Large Language Models (LLMs): Mistral3.1\footnote{\url{https://hf.co/RedHatAI/Mistral-Small-3.1-24B-Instruct-2503-FP8-dynamic}}, DeepHermes\footnote{\url{https://hf.co/NousResearch/DeepHermes-3-Mistral-24B-Preview}}, and Qwen3\footnote{\url{https://hf.co/Qwen/Qwen3-30B-A3B-FP8}}.
Mistral3.1 and DeepHermes are based on the Mistral 24B architecture, while Qwen3 is a 30B parameter mixture-of-experts model optimized for reasoning.
All models are deployed with FP8 quantization via vLLM \cite{kwon2023efficient} on a single GPU (NVIDIA L40S, 48\,GB VRAM), balancing resource constraints and usability.
Sampling parameters follow the respective authors' recommendations: max\_tokens=4096 for all models, temperature=0.15 for Mistral3.1 and DeepHermes, temperature=0.7 with top\_p=0.8 for Qwen3, and temperature=0.6 with top\_p=0.95 for Qwen3 with reasoning.
Constrained decoding with a predefined JSON schema is activated to ensure structured outputs.

\subsection{Prompt Strategies}\label{sc:prompts}
\begin{table}[!th]
\centering
\caption{\label{tab:text}Linguistic, speech, cognitive, and emotional features for spoken language transcripts (T).}
\begin{tabular}{c||l|l}
\toprule
\textbf{T}  & \textbf{category}          & \textbf{question}                           \\
\midrule
\textbf{1}  &  Expressiveness            & How expressive is the language?            \\
\textbf{2}  &  Lexical Diversity         & How diverse is the vocabulary?                  \\
\textbf{3}  &  Grammatical Errors        & How incorrect is the grammar?                  \\
\textbf{4}  &  Coherence                 & How logical are statements connected?          \\
\textbf{5}  &  Cohesion                  & How well are text elements connected?          \\
\textbf{6}  &  Fluency                   & How fluent is the speech?                      \\
\textbf{7}  &  Precision                 & How precise are the statements?                \\
\textbf{8}  &  Clarity                   & How easy is the language to understand?        \\
\textbf{9}  &  Sentence Length           & How complex are the sentences?                 \\
\textbf{10} &  Pauses                    & Are there pauses, filler words, hesitations? \\
\textbf{11} &  Speech rate               & How fast is the speech rate?                   \\
\textbf{12} &  Confabulations            & Is there invented content?                     \\
\textbf{13} &  Paraphrases               & Are there incorrectly used statements?         \\
\textbf{14} &  Word-finding              & Are there difficulties in finding words?       \\
\textbf{15} &  Perplexity                & Are there confusing statements?              \\
\textbf{16} &  Redundancy                & Are there repetitive statements?               \\
\textbf{17} &  Cognitive Flexibility     & How flexible is the thinking process?          \\
\textbf{18} &  Emotional Tone            & How pronounced is the emotional tone?          \\
\textbf{19} &  Lack of Drive             & Does the language lack in energy?      \\
\textbf{20} &  Negativity                & How pessimistic is the language?      
\\
\bottomrule
\end{tabular}
\end{table}
All prompts are formulated in German to match the input language; English translations are provided here for reporting purposes. 
Original prompts will be released on GitHub upon acceptance.

A consistent \textbf{system prompt} is used across all models and experiments:
\begin{tcolorbox}[colframe=black!60, colback=gray!10, boxrule=0.5pt, width=\columnwidth, boxsep=1pt,  left=2pt, right=2pt, top=2pt, bottom=2pt]
\textit{``You are an expert in assessing speech intelligibility in the medical field. 
Your task is to identify signs of dementia and depression in people based on transcribed conversations. ... '' + JSON schema}
\end{tcolorbox}
For ground truth transcriptions, the system prompt includes the transcription rules; for pause-enriched transcripts, the pause notation definition is added; for reasoning-enabled settings, a deep-thinking prompt is appended.
All prompts conclude with an instruction to produce output according to a task-specific JSON schema (cf.\ Sec.~\ref{sc:llm}).

To introduce the conversational and clinical context, a \textbf{task-specific user prompt} is included at the beginning of all prompts for feature extraction and zero-shot prediction:
\begin{tcolorbox}[colframe=black!60, colback=gray!10, boxrule=0.5pt, width=\columnwidth, boxsep=1pt,  left=2pt, right=2pt, top=2pt, bottom=2pt] 
\textit{``You will be given the transcript of a neuropsychological history taking interview between Person A and an Interviewer. 
This includes questions about memory, attention, daily activities, orientation, mood, sleep, appetite, physical activity, and medication.''}
\end{tcolorbox}
We evaluate two prompting strategies: (1) a \textit{joined} strategy, where feature extraction and zero-shot prediction are combined within a single user prompt, and (2) a \textit{sequential} strategy, where both tasks are prompted in separate turns while retaining the full chat history.
This comparison evaluates trade-offs between integrated context reasoning (\textit{joined}) and task-specific focus (\textit{sequential}) for clinical assessment.

To leverage LLMs to directly predict the GDS and GDS-D, each task-specific user prompt is extended by the \textbf{zero-shot prediction prompt}, followed by the definitions of both scales:
\begin{tcolorbox}[colframe=black!60, colback=gray!10, boxrule=0.5pt, width=\columnwidth, boxsep=1pt,  left=2pt, right=2pt, top=2pt, bottom=2pt] 
\textit{``Task: Your task is to rate Person A on a scale for dementia (0.0--4.0) and depression (0.0--4.0) as follows:'' + definitions}
\end{tcolorbox} 
The model is instructed to output floating-point values from 0.0 to 4.0 (corresponding to stages 1--5), as labels may fall between transition stages.
The last two stages are excluded, as severe levels are not represented in the dataset.

To leverage LLMs to extract predefined feature sets, the task-specific user prompt is extended by the \textbf{feature extraction prompt} as follows:
\begin{tcolorbox}[colframe=black!60, colback=gray!10, boxrule=0.5pt, width=\columnwidth, boxsep=1pt, left=2pt, right=2pt, top=2pt, bottom=2pt]
\textit{``Task: Your task is to rate the speech of Person A about \{N\} features with a score between 0.0 and 1.0, where 0.0 stands for 'not present' and 1.0 for 'strongly pronounced'.''}
\end{tcolorbox} 
where \textit{N} denotes the size of the respective feature set.
The model outputs floating-point values from 0.0 to 1.0.
Each feature set is specified in the task-specific user prompt and includes features derived from:
\begin{itemize}
    \item \textbf{Symptoms (N=8):} Reported symptoms in standardized neuropsychological interviews (Tab.~\ref{tab:infr}).
    \item \textbf{Observations (N=10):} Behavioral \& psychopathologic observations in standardized neuropsychological interviews (Tab.~\ref{tab:infr}).
    \item \textbf{Conversation (N=9):} Conversational structure and verbal behavior as defined by the transcription rules of Dresing \& Pehl \cite{dresing_pehl10}.
    \item \textbf{Language (N=20):} Text, speech, cognitive, and emotional markers in spoken language transcripts (Tab.~\ref{tab:text}).
\end{itemize}

\section{Experiments}
All experiments use human and pause-enriched ASR transcripts from history taking interviews, with both \textit{joined} and \textit{sequential} prompting strategies (Sec.~\ref{sc:llm},~\ref{sc:prompts}).
For the prediction of GDS and GDS-D scores, we compare Mistral3.1, DeepHermes, and Qwen3 in two settings: (1) \textbf{zero-shot prediction} and (2) \textbf{feature extraction} for downstream regression modeling.
For the feature extraction setting, we train Support Vector Regression (SVR) models with Linear and Radial Basis Function (RBF) kernels to predict GDS and GDS-D severity scores.
We employ stratified five-fold cross-validation (5-fold CV) with speaker-disjoint training (${\sim}80\%$) and test (${\sim}20\%$) splits.
Optimal SVR hyperparameters are determined via nested grid search on the training portion, with $\gamma \in \{10^{-k} \mid k = 1, \ldots, 5 \}$ and $C \in \{10^{k} \mid k = -1, \ldots, 2 \}$.
Performance is evaluated using the mean absolute error (MAE), which directly quantifies the average deviation from expert-rated severity scores in interpretable scale units.

\section{Results}
\begin{table}[!th]
\centering
\caption{\label{tab:zsdem}Mean absolute error for \textbf{zero-shot prediction} of \textbf{GDS (dementia)} and \textbf{GDS-D (depression)}, standalone (alone) and after feature extraction per feature set. Values: \textit{sequential}/\textit{joined} prompting. GT~=~human ground truth, ASR~=~pause-enriched whisper transcriptions, +think~=~reasoning mode.}

\begin{tabular}{c|l||c|c|c|c|c}
\toprule
 & \textbf{Model} & \textbf{alone} & \textbf{Conversation} & \textbf{Symptoms} & \textbf{Observations} & \textbf{Language} \\ 
\midrule
\multicolumn{7}{c}{\textbf{GDS (dementia)}} \\
\midrule
\multirow{5}{*}{\rotatebox[origin=c]{90}{\begin{tabular}{c} \textbf{GT} \end{tabular}}}
 & \textbf{Mistral} & 1.29 & 1.25/1.20 & 1.25/1.27 & 1.24/1.30 & \textbf{1.19}/1.26 \\
 & \textbf{Hermes} & 1.35 & 1.31/1.51 & 1.31/1.51 & 1.42/1.65 & \textbf{1.25}/1.44 \\
 & \textbf{+think} & 1.35 & 1.30/1.43 & 1.31/1.48 & 1.34/1.54 & \textbf{1.23}/1.40 \\
 & \textbf{Qwen} & 1.46 & 1.41/1.55 & 1.45/1.46 & 1.36/1.42 & \textbf{1.34}/1.50 \\
 & \textbf{+think} & 1.56 & 1.68/1.64 & 1.49/1.55 & 1.46/1.57 & \textbf{1.39}/1.51 \\

\midrule

\multirow{5}{*}{\rotatebox[origin=c]{90}{\begin{tabular}{c} \textbf{ASR} \end{tabular}}}
 & \textbf{Mistral} & 1.17 & 1.24/1.19 & 1.24/1.23 & 1.23/1.19 & 1.13/\textbf{1.11} \\
 & \textbf{Hermes} & 1.20 & 1.21/1.30 & 1.28/1.43 & 1.23/1.32 & \textbf{1.16}/1.21 \\
 & \textbf{+think} & 1.18 & 1.20/1.21 & 1.22/1.36 & 1.21/1.27 & \textbf{1.16}/1.21 \\
 & \textbf{Qwen} & 1.39 & 1.29/1.35 & 1.33/1.40 & 1.35/1.33 & \textbf{1.26}/1.30 \\
 & \textbf{+think} & 1.41 & 1.41/1.39 & 1.46/1.46 & 1.48/1.49 & \textbf{1.28}/1.35 \\

 \midrule
 \multicolumn{7}{c}{\textbf{GDS-D (depression)}} \\
 \midrule

 \multirow{5}{*}{\rotatebox[origin=c]{90}{\begin{tabular}{c} \textbf{GT} \end{tabular}}}
 & \textbf{Mistral} & \textbf{0.61} & 0.64/0.63 & 0.66/0.62 & 0.65/0.62 & 0.62/0.64 \\
 & \textbf{Hermes} & 0.66 & 0.61/0.69 & 0.64/0.69 & 0.65/\textbf{0.60} & 0.68/0.63 \\
 & \textbf{+think} & 0.63 & 0.64/0.64 & 0.62/0.64 & 0.63/\textbf{0.61} & 0.65/0.63 \\
 & \textbf{Qwen} & 0.65 & 0.68/0.65 & 0.64/0.63 & 0.70/0.69 & 0.66/\textbf{0.61} \\
 & \textbf{+think} & 0.69 & 0.76/0.69 & \textbf{0.61}/0.66 & 0.68/0.67 & 0.65/0.62 \\

\midrule

\multirow{5}{*}{\rotatebox[origin=c]{90}{\begin{tabular}{c} \textbf{ASR} \end{tabular}}}
 & \textbf{Mistral3.1} & \textbf{0.60} & 0.60/0.60 & 0.70/0.62 & 0.65/0.65 & 0.67/0.65 \\
 & \textbf{DeepHermes} & \textbf{0.60} & 0.68/0.65 & 0.71/0.68 & 0.67/0.66 & 0.72/0.68 \\
 & \textbf{+think} & 0.64 & 0.69/\textbf{0.63} & 0.68/0.66 & 0.67/0.65 & 0.70/0.66 \\
 & \textbf{Qwen3} & 0.66 & 0.67/0.68 & 0.66/\textbf{0.60} & 0.75/0.71 & 0.71/0.69 \\
 & \textbf{+think} & 0.64 & 0.71/0.72 & 0.68/\textbf{0.63} & 0.70/0.64 & 0.68/0.67 \\

\bottomrule
\end{tabular}
\end{table}

\begin{table}[!th]
\centering
\caption{\label{tab:ftdem}Mean absolute error and standard deviation (5-fold cross-validation) for \textbf{LLM feature extraction + SVR prediction} of \textbf{GDS (dementia)} and \textbf{GDS-D (depression)} per feature set. Notation as in Tab.~\ref{tab:zsdem}.}

\begin{tabular}{c|l||c|c|c|c}
\toprule
 & \textbf{Model} & \textbf{Conversation} & \textbf{Symptoms} & \textbf{Observations} & \textbf{Language} \\
\midrule
\multicolumn{6}{c}{\textbf{GDS (dementia)}} \\
\midrule

\multirow{5}{*}{\rotatebox[origin=c]{90}{\begin{tabular}{c} \textbf{GT} \end{tabular}}}
& \textbf{Mistral} & .94$\pm$.12/.86$\pm$.11 & \textbf{.81$\pm$.11}/.83$\pm$.11 & .86$\pm$.18/.83$\pm$.11 & .89$\pm$.17/.82$\pm$.18 \\
& \textbf{Hermes} & .93$\pm$.11/.89$\pm$.13 & \textbf{.85$\pm$.12}/\textbf{.85$\pm$.14} & .91$\pm$.12/.87$\pm$.10 & .91$\pm$.19/.89$\pm$.17 \\
& \textbf{+think} & .96$\pm$.08/.92$\pm$.12 & .86$\pm$.16/.87$\pm$.14 & \textbf{.85$\pm$.10}/.93$\pm$.15 & \textbf{.85$\pm$.14}/\textbf{.85$\pm$.11} \\
& \textbf{Qwen} & .91$\pm$.10/.89$\pm$.07 & .84$\pm$.16/\textbf{.81$\pm$.12} & .82$\pm$.19/.84$\pm$.19 & \textbf{.81$\pm$.13}/\textbf{.81$\pm$.12} \\
& \textbf{+think} & .94$\pm$.10/.89$\pm$.09 & .82$\pm$.14/.86$\pm$.14 & .88$\pm$.16/.89$\pm$.12 & .88$\pm$.15/\textbf{.78$\pm$.11} \\

\midrule

\multirow{5}{*}{\rotatebox[origin=c]{90}{\begin{tabular}{c} \textbf{ASR} \end{tabular}}}
& \textbf{Mistral} & .92$\pm$.15/.83$\pm$.13 & .89$\pm$.13/.87$\pm$.12 & .86$\pm$.11/.88$\pm$.10 & \textbf{.80$\pm$.11}/\textbf{.80$\pm$.15} \\
& \textbf{Hermes} & .95$\pm$.14/.95$\pm$.14 & .82$\pm$.10/\textbf{.80$\pm$.13} & .88$\pm$.11/.88$\pm$.11 & .87$\pm$.13/.84$\pm$.10 \\
& \textbf{+think} & .98$\pm$.18/.96$\pm$.13 & \textbf{.81$\pm$.10}/.84$\pm$.16 & .90$\pm$.12/.92$\pm$.10 & \textbf{.81$\pm$.10}/.85$\pm$.17 \\
& \textbf{Qwen} & .95$\pm$.11/.82$\pm$.10 & .86$\pm$.15/.89$\pm$.15 & .85$\pm$.08/.86$\pm$.10 & .85$\pm$.19/\textbf{.82$\pm$.16} \\
& \textbf{+think} & .91$\pm$.11/.90$\pm$.14 & .90$\pm$.11/\textbf{.81$\pm$.14} & .90$\pm$.15/.87$\pm$.11 & .84$\pm$.14/.86$\pm$.15 \\

\midrule
\multicolumn{6}{c}{\textbf{GDS-D (depression)}} \\
\midrule

\multirow{5}{*}{\rotatebox[origin=c]{90}{\begin{tabular}{c} \textbf{GT} \end{tabular}}}
& \textbf{Mistral} & .97$\pm$.10/.93$\pm$.08 & \textbf{.58$\pm$.05}/.62$\pm$.04 & .63$\pm$.07/.63$\pm$.08 & .63$\pm$.06/.62$\pm$.04 \\
& \textbf{Hermes} & .94$\pm$.10/.95$\pm$.10 & \textbf{.61$\pm$.07}/.67$\pm$.10 & .67$\pm$.04/.64$\pm$.03 & .68$\pm$.06/.65$\pm$.10 \\
& \textbf{+think} & 1.01$\pm$.08/.93$\pm$.07 & \textbf{.63$\pm$.06}/.67$\pm$.11 & .67$\pm$.06/.66$\pm$.07 & .71$\pm$.07/.67$\pm$.07 \\
& \textbf{Qwen} & 1.00$\pm$.07/.94$\pm$.09 & \textbf{.67$\pm$.08}/.65$\pm$.09 & .73$\pm$.07/.70$\pm$.05 & .69$\pm$.06/\textbf{.67$\pm$.04} \\
& \textbf{+think} & .96$\pm$.07/.89$\pm$.14 & .76$\pm$.10/\textbf{.65$\pm$.03} & .70$\pm$.07/.67$\pm$.07 & .68$\pm$.09/.67$\pm$.06 \\

\midrule

\multirow{5}{*}{\rotatebox[origin=c]{90}{\begin{tabular}{c} \textbf{ASR} \end{tabular}}}
& \textbf{Mistral} & .90$\pm$.08/.96$\pm$.16 & \textbf{.60$\pm$.08}/.66$\pm$.06 & .62$\pm$.07/.73$\pm$.08 & .69$\pm$.08/.70$\pm$.11 \\
& \textbf{Hermes} & .97$\pm$.12/.88$\pm$.14 & .72$\pm$.05/.76$\pm$.10 & .68$\pm$.05/.73$\pm$.06 & .70$\pm$.03/\textbf{.67$\pm$.11} \\
& \textbf{+think} & .89$\pm$.10/.90$\pm$.10 & .68$\pm$.07/.78$\pm$.04 & .75$\pm$.07/.70$\pm$.06 & .70$\pm$.08/\textbf{.67$\pm$.10} \\
& \textbf{Qwen} & .86$\pm$.08/.88$\pm$.04 & .70$\pm$.06/\textbf{.64$\pm$.11} & .73$\pm$.08/.72$\pm$.04 & .72$\pm$.05/.75$\pm$.14 \\
& \textbf{+think} & .97$\pm$.05/.94$\pm$.14 & .79$\pm$.11/.75$\pm$.16 & .76$\pm$.08/.81$\pm$.15 & \textbf{.64$\pm$.09}/.67$\pm$.11 \\

\bottomrule
\end{tabular}
\end{table}

Tab.~\ref{tab:zsdem} presents the MAE for zero-shot prediction of the GDS and GDS-D across LLMs, transcription types, and prompting strategies.
For GDS (dementia), preceding zero-shot prediction with language feature extraction in a sequential prompt consistently yields the best results, likely reflecting the prominence of linguistic markers in cognitive decline.
Pause-enriched ASR transcripts outperform human ground truth in most conditions (e.g., Mistral~3.1: 1.17 vs.\ 1.29 for zero-shot only), suggesting that temporal pause information compensates for missing conversational structure markers.
Mistral~3.1 achieves the overall lowest GDS MAE of 1.11 using language features on pause-enriched transcripts.
For GDS-D (depression), MAEs are substantially lower (0.60--0.69 vs.\ 1.11--1.56 for GDS), indicating that affective symptom patterns are more directly accessible from interview language.
In contrast to GDS, prior feature extraction does not consistently improve depression zero-shot predictions; the zero-shot-only setting often achieves the best results (e.g., Mistral~3.1: 0.60 on pause-enriched transcripts).
Across both tasks, reasoning-enhanced configurations yield mixed results.

Tab.~\ref{tab:ftdem} reports MAE and standard deviation (SD) (5-fold CV) for SVR models trained on LLM-extracted feature sets.
For dementia, MAEs improve substantially over zero-shot baselines: language features yield the lowest MAEs of 0.78 (Qwen3+reasoning, ground truth) and 0.80 (Mistral3.1, pause-enriched ASR), while symptom features achieve competitive results despite smaller dimensionality.
Conversation features perform moderately for dementia but are notably less effective for depression (MAEs of 0.86--1.01), suggesting that structural markers are more informative for cognitive than affective assessment.
For depression, symptom features in the sequential setting yield the best results, with Mistral~3.1 achieving MAEs of 0.58 (ground truth) and 0.60 (pause-enriched ASR) showing only marginal improvements over zero-shot predictions.
Structured symptom ratings particularly benefit from conversational annotations in ground truth transcripts, while pause-enriched ASR transcripts additionally leverage language features.
Across both tasks, Mistral~3.1 outperforms other models in most conditions.
Combining the $n$-best feature sets and zero-shot predictions in SVR models (reported for Mistral3.1, pause-enriched ASR only), dementia prediction reaches MAE of 0.76$\pm$0.15, while for depression no combination outperforms the best individual feature set.

\section{Conclusion}
We investigated open-weights LLMs for automated dementia and depression assessment from clinical history taking interviews.
To this end, we introduced an observer-based Global Depression Scale aligned with the Global Deterioration Scale, enabling parallel staging of affective and cognitive symptoms.
Our results demonstrate that LLMs effectively predict depression severity in zero-shot settings, while dementia assessment requires explicit feature extraction and regression modeling to achieve comparable performance.
Model errors remained below 1.0 MAE in most configurations, indicating high scoring accuracy relative to expert ratings.
Structured symptom ratings, language features, and pause-enriched ASR transcripts further improved prediction performance.
These findings highlight the potential of LLM-based zero-shot screening for depression and the value of structured feature extraction for dementia staging in the context of differential diagnostics.
Future work should explore multimodal and fine-tuned LLMs for enhanced neuropsychiatric assessment.

\subsubsection*{Acknowledgments.}
Funded by the Deutsche Forschungsgemeinschaft (DFG) -- Project Number 549142762 -- FIP 160.

\bibliographystyle{splncs04}
\bibliography{tsd1364a}

\end{document}